\documentclass[twocolumn,american,pre,showpacs,preprintnumbers,amsmath,amssymb]{revtex4-1}
\usepackage[latin9]{inputenc}
\usepackage{amsbsy}
\usepackage{amstext}
\usepackage{amssymb}
\usepackage{graphicx}
\usepackage{esint}

\makeatletter

\usepackage{dcolumn}
\usepackage{bm}

\usepackage{babel}

\usepackage{babel}

\makeatother

\usepackage{babel}
\begin{document}
\global\long\def\bea{ 
\begin{eqnarray}
\end{eqnarray}
 }
 \global\long\def\eea{ {eqnarray}}
 \global\long\def\bit{\begin{itemize}\end{itemize}}
 \global\long\def\eit{ {itemize}}

\global\long\def\be{ 
\begin{equation}
\end{equation}
 }
 \global\long\def\ee{{equation}}
 \global\long\def\ra{\rangle}
 \global\long\def\la{\langle}
 \global\long\def\U{\widetilde{U}}


\global\long\def\bra#1{{\langle#1|}}
 \global\long\def\ket#1{{|#1\rangle}}
 \global\long\def\bracket#1#2{{\langle#1|#2\rangle}}
 \global\long\def\inner#1#2{{\langle#1|#2\rangle}}
 \global\long\def\expect#1{{\langle#1\rangle}}
 \global\long\def\e{{\rm e}}
 \global\long\def\proj{{\hat{{\cal P}}}}
 \global\long\def\tr{{\rm Tr}}
 \global\long\def\H{{\hat{H}}}
 \global\long\def\Hdag{{\hat{H}}^{\dagger}}
 \global\long\def\Lop{{\cal L}}
 \global\long\def\Ehat{{\hat{E}}}
 \global\long\def\Edag{{\hat{E}}^{\dagger}}
 \global\long\def\Shat{\hat{S}}
 \global\long\def\Sdag{{\hat{S}}^{\dagger}}
 \global\long\def\Ahat{{\hat{A}}}
 \global\long\def\Adag{{\hat{A}}^{\dagger}}
 \global\long\def\U{{\hat{U}}}
 \global\long\def\Udag{{\hat{U}}^{\dagger}}
 \global\long\def\Zhat{{\hat{Z}}}
 \global\long\def\Phat{{\hat{P}}}
 \global\long\def\Op{{\hat{O}}}
 \global\long\def\id{{\hat{I}}}
 \global\long\def\x{{\hat{x}}}
 \global\long\def\P{{\hat{P}}}
 \global\long\def\Px{\proj_{x}}
 \global\long\def\Pr{\proj_{R}}
 \global\long\def\Pl{\proj_{L}}


\title{OTOC, complexity and entropy in bi-partite systems}

\author{Pablo D. Bergamasco}

\affiliation{Departamento de Física, CNEA, Libertador 8250, (C1429BNP) Buenos 
Aires, Argentina}

\author{Gabriel G. Carlo}

\affiliation{Departamento de Física, CNEA, CONICET, Libertador 8250, (C1429BNP) Buenos
Aires, Argentina}

\author{Alejandro M. F. Rivas}

\affiliation{Departamento de Física, CNEA, CONICET, Libertador 8250, (C1429BNP) Buenos
Aires, Argentina}

\email{pablobergamasco@cnea.gov.ar,carlo@tandar.cnea.gov.ar,rivas@tandar.cnea.gov.ar}

\selectlanguage{american}%

\date{\today}

\begin{abstract}
There is a remarkable interest in the study of Out-of-time ordered correlators (OTOCs) that goes from 
many body theory and high energy physics to quantum chaos. In this latter case there is a special focus 
on the comparison with the traditional measures of quantum complexity such as the spectral statistics, for example. 
The exponential growth has been verified for many paradigmatic maps 
and systems. But less is known for multi-partite cases. On the other hand the recently introduced 
Wigner separability entropy (WSE) and its classical counterpart (CSE) provide with a complexity measure 
that treats equally quantum and classical distributions in phase space. We have compared the behavior of 
these measures in a system consisting of two coupled and perturbed cat maps with different dynamics: 
double hyperbolic (HH), double elliptic (EE) and mixed (HE). 
In all cases, we have found that the OTOCs and the WSE have essentially the same behavior, providing 
with a complete characterization in generic bi-partite systems and at the same time 
revealing them as very good measures of quantum complexity for phase space distributions. 
Moreover, we establish a relation between both quantities by means of a recently proven theorem 
linking the second Renyi entropy and OTOCs.
\end{abstract}

\pacs{05.45.Mt, 05.45.Pq, 03.67.Mn, 03.65.Ud}

\maketitle

{\em Introduction.}--Nowadays there is a huge interest in OTOCs among the quantum chaos community, where they have been related to 
traditional measures as spectral statistics and the like \cite{Qchaos}. These correlators were first introduced 
in superconductivity studies \cite{Larkin}, where their exponential growth over time was associated to chaotic behavior. 
As a matter of fact, if we adopt the usual definition of the OTOC given by 
\begin{equation}
C(t) = \left\langle [\hat{A}(t),B][\hat{A}(t),B]^{\dagger} \right\rangle, 
\label{eq:otoc}
\end{equation} 
(i.e., the thermal average $\left\langle \cdot \right\rangle= {\rm Tr}(\cdot)/N$ 
of the commutator between two operators at different times, with $N$ the Hilbert space dimension), 
and we take $\hat{A}=\hat{X}$ and $\hat{B}=\hat{P}$ as the position and 
momentum operators respectively, it is easy to show this. The commutator leads 
to the Poisson bracket at the semiclassical level, which in turn grows exponentially 
at a rate of twice the Lyapunov exponent. In \cite{Garcia}, this exponential growth for 
the (one dimensional) quantum perturbed Arnold cat map has been shown. 
Also, in \cite{Lakshminarayan} a growth at half this rate was discovered in the baker map, using 
projectors instead of position and momentum operators. Finally, there was some controversy around 
the Bunimovich stadium case in \cite{Hashimoto}, which has been explained by means of 
replacing the thermal average with Gaussian minimal uncertainty wavepackets (the ``most classical'' initial 
state) \cite{Rozenbaum}.

But previously the surge in interest came from the many body area \cite{Shenker,Aleiner,Huang,Borgonovi}. 
As a prominent feature, an upper bound for the OTOC growth was conjectured in the context 
of black hole models \cite{Maldacena}. Also, the OTOCs have been related to multiple quantum coherences and 
used as an entanglement witness in NMR \cite{Gaertner}. A path integral approach has been presented, in which 
the OTOCs are expressed as coherent sums over contributions from different mean field solutions \cite{Rammensee}. 
More recently, the OTOC behavior has been determined 
for one of the simplest examples of multi-particle systems. This corresponds to a bi-partite system consisting of 
two strongly chaotic and weakly coupled kicked rotors \cite{Lakshminarayan2}. It was found that the scrambling 
process has two phases, one in which the exponential growth is 
intra subsystem and a second one which only depends on the interaction.  

Recently, in the spirit of algorithmic complexity, the Wigner separability entropy (WSE) \cite{Benenti} 
and the Classical Separability Entropy (CSE) \cite{Prosen} have been introduced as measures of complexity that put 
quantum and classical distributions (in phase space) on an equal footing. We have characterized how the WSE and the CSE behave for a 
bi-partite system given by two coupled and perturbed cat maps with different dynamics. Three cases were considered, one 
where both maps are hyperbolic (chaotic) (HH), one where both are elliptic (regular) (EE), and a mixed situation 
where one map is hyperbolic and the other elliptic (HE) \cite{Bergamasco}. In this work we have set a twofold objective, 
on one hand we have investigated the behavior of OTOCs for these different scenarios, and on the other hand we have 
compared them with the complexity measures previously mentioned. As a result, we have found that the OTOCs are in general 
good indicators of complexity that can be related to WSE and CSE via the so called OTOC-RE theorem \cite{Hosur,Fan}. 

{\em OTOCs and WSE.}--In Eq. \ref{eq:otoc} we have defined these correlators in the most commonly found way, i.e. by performing 
a thermal average of the commutator of two operators, one of which evolves with time in a Heisenberg fashion. 
For our purposes which are investigating the behavior for different kinds of dynamics in each subsystem it is 
preferable to calculate the expectation value on a given initial state, much in the same way as we have 
done previously to compute the phase space entropies WSE and CSE \cite{Bergamasco}. This is accomplished by  
$\left\langle\cdot\right\rangle = {\rm Tr}(\rho \cdot)$, where $\rho$ is the density matrix of a ``classical like'' 
initial state, which we take to be a coherent state. 

Also, there is freedom in the choice of operators $\hat{A}$ and $\hat{B}$. As mentioned, we can take 
$\hat{X}$ and $\hat{P}$ in order to formally associate them to a Lyapunov exponential 
growth. But we will also consider $\hat{B}=\hat{\rho}(0)$, the density operator of the initial state. 
The cat map is defined on the torus and we use an approximation to position and momentum operators 
in the classical limit that makes use of the Schwinger shift operators \cite{Schwinger} 
\begin{equation}
	\hat{V} = \sum_{q\in \mathbb{Z_{N}}}{\ket{q+1}\bra{q}}\ ;
	\ \hat{U} = \sum_{q\in \mathbb{Z_{N}}}{\ket{q}\bra{q}\tau^{2q}}
\label{eq:shiftop}
\end{equation} 
with $\tau = \exp{i\pi/N}$. Position and momentum operators for each one degree of freedom map can be written as
\begin{equation}
	\hat{X} = \frac{\hat{U}-\hat{U}^{\dagger}}{2i}\ ;\ \hat{P} = \frac{\hat{V}-\hat{V}^{\dagger}}{2i}. 
\label{eq:XP1D}
\end{equation}
These operators are readily extended to the two degrees of freedom bi-partite space as the tensor product of 
similar operators acting on each subsystem (labeled as $1$ and $2$), 
\begin{equation}
\hat{X}^{\rm 2D} = \hat{X}^{1}\otimes\hat{X}^{2}\ ;\ \hat{P}^{\rm 2D} = \hat{P}^{1}\otimes\hat{P}^{2}.
\label{eq:XP2D}
\end{equation} 
It is worth mentioning that when the operators $\hat{A}$ and $\hat{B}$ are Hermitian the OTOC of Eq. \ref{eq:otoc} 
can be expressed as the correlators difference: 
\begin{equation}
C(t) = -2 [C_4(t)-C_2(t)]/N,  
\label{eq:otocsplit}
\end{equation} 
where $C_2(t)={\rm Tr}[\hat{A}^2(t) \hat{B}^2]$ (a 2 point correlator), and 
$C_4(t)={\rm Tr}[\hat{A}(t) \hat{B} \hat{A}(t) \hat{B}]$ (a 4 point correlator). 
Also, our ${\rm Tr}(\rho \cdot)$ operation when $\hat{B}=\hat{\rho}(0)$ corresponds to a pure state, 
can be proven to be equivalent to ${\rm Tr}(\cdot)/2$, making it similar to the thermal average times $N$.

We now briefly explain the WSE and CSE definitions and main properties. 
A good analogue of Liouville distributions in classical mechanics is given by the Wigner distributions in 
phase space $W(\boldsymbol{x})$, which are defined in terms of the Weyl-Wigner symbol of the density operator as
\begin{equation}
W(\boldsymbol{x})=(2\pi\hbar)^{-d/2}\rho(\boldsymbol{x})=(2\pi\hbar)^{-d/2}
\text{Tr}\left[\hat{R}_{\boldsymbol{x}}\hat{\rho}\right], 
\end{equation}
where $\hat{R}_{\boldsymbol{x}}$ forms a basis set of unitary reflection operators
on points $\boldsymbol{x}\equiv(\boldsymbol{q},\boldsymbol{p})$ \cite{ozrep,opetor}.
The Schmidt decomposition of the density operator is given by 
\begin{equation}
\hat{\rho}=\sum\sigma_{n}\hat{a}_{n}\otimes\hat{b}_{n},
\label{SVDrho}
\end{equation}
where $\{\hat{a}_{n}\}$ and $\{\hat{b}_{n}\}$ are orthonormal
bases for the Hilbert-Schmidt operator spaces $B(\mathcal{H}_{1})$ and $B(\mathcal{H}_{2})$, 
respectively. This directly leads to the Schmidt (singular value) decomposition of the Wigner function given by 
\begin{equation}
W(\boldsymbol{x})=\sum_{n}\sigma_{n}a_{n}(\boldsymbol{x_1})b_{n}(\boldsymbol{x_2}),
\label{eq:SVDWigner}
\end{equation}
where $\{a_{n}\}$ and $\{b_{n}\}$ are now orthonormal bases for
$L^{2}(\Omega_{1})$ and $L^{2}(\Omega_{2})$ 
(which are associated to the Hilbert space decomposition), such that:
\[
a_{n}(\boldsymbol{x_1})=\text{Tr}\left[\hat{R}_{\boldsymbol{x_1}}\hat{a}_{n}\right],
\quad\textrm{and}\quad b_{n}(\boldsymbol{x_2})=\text{Tr}\left[\hat{R}_{\boldsymbol{x_2}}\hat{b}_{n}\right].
\]
The WSE is defined in \cite{Benenti} as
$
h[W]=-\sum_{n}\tilde{\sigma}_{n}^{2}\ln\tilde{\sigma}_{n}^{2}
$, 
where 
$
\tilde{\sigma}_{n}\equiv{\sigma_{n}}/{\sqrt{\int d\boldsymbol{x}W^{2}(\boldsymbol{x})}}.
$

The WSE $h[W]$ provides a measure of separability of the Wigner function with respect to a given 
phase space decomposition. One of the main properties of the WSE is that its classical analogue 
is the CSE (or s-entropy) $h[\rho_{c}]$ defined in \cite{Prosen}, where a discretized classical phase 
space distribution is used instead of the Wigner function $W(\boldmath{x})$. This makes 
the complexity concept in both the quantum and the classical world fully equivalent. It is very important to mention that 
for a pure state $\hat{\rho}=|\psi\rangle\langle\psi|$, $h[W]=-2S_{\rm VN}(\hat{\rho}_{1})=-2S_{\rm VN}(\hat{\rho}_{2})$, 
where $\hat{\rho}_{1}=\text{Tr}_{2}(\hat{\rho})$ and $\hat{\rho}_{2}=\text{Tr}_{1}(\hat{\rho})$
are the reduced density operators for subsystems $1$ and $2$, and $S_{\rm VN}$
is the von Neumann entropy.

{\em Model system.}--The quantum cat map \cite{Hannay 1980} 
is paradigmatic in the quantum chaos area 
\cite{Hannay 1980,Ozorio 1994,Haake,Espositi 2005}.
We consider the behavior of two coupled
perturbed cat maps, a two degrees of freedom example. 
These two maps can have different types of dynamics.

Each degree of freedom is defined on the 2-torus as \cite{Hannay 1980}
\begin{equation}
\left(\begin{array}{c}
q_{t+1}\\
p_{t+1}
\end{array}\right)=\mathcal{M}\left(\begin{array}{c}
q_{t}\\
p_{t}+\epsilon\left(q_{t}\right)
\end{array}\right)
\end{equation}
with $q$ and $p$ taken modulo $1$, and 
\[
\epsilon\left(q_{t}\right)=-\frac{K}{2\pi}\sin\left(2\pi q_{t}\right). 
\]
For the ergodic case we have chosen the hyperbolic map 
\begin{equation}
\mathcal{M}_{h}=\left(\begin{array}{cc}
2 & 1\\
3 & 2
\end{array}\right),\label{eq:Mhyper}
\end{equation}
while for the regular behavior we use the elliptic map 
\begin{equation}
\mathcal{M}_{e}=\left(\begin{array}{cc}
0 & 1\\
-1 & 0
\end{array}\right).\label{eq:Meliptic}
\end{equation}
Torus quantization amounts to have a finite Hilbert space of
dimension $N=\frac{1}{2\pi\hbar}$, with discrete positions and momenta 
in a lattice of separation $\frac{1}{N}$
\cite{Hannay 1980}. In coordinate representation the propagator is given by 
a $N\times N$ unitary matrix  
\begin{equation}
U_{jk}=A{\exp}\left[\frac{i\pi}{N\mathcal{M}_{12}}(\mathcal{M}_{11}j^{2}-2jk
+\mathcal{M}_{22}k^{2})+F\right],\label{uqq}
\end{equation}
where 
$
A=[1/\left(iN\mathcal{M}_{12}\right)]^{1/2}  
$, and 
$
F=[iKN/(2\pi)]\cos(2\pi j/N)
$.
The states $\langle q|\mathbf{q}_{j}\rangle$ are periodic 
Dirac delta distributions at positions $q=j/N{\rm mod}(1)$, with $j$
integer in $[0,N-1]$.

The two degrees of freedom classical system is defined in a four-dimensional 
phase space of coordinates $\left(q^{1},q^{2},p^{1},p^{2}\right)$ 
\cite{Benenti} as 
\[
\left(\begin{array}{c}
q_{t+1}^{1}\\
p_{t+1}^{1}
\end{array}\right)=\mathcal{M}_{1}\left(\begin{array}{c}
q_{t}^{1}\\
p_{t}^{1}+\epsilon\left(q_{t}^{1}\right)+\kappa\left(q_{t}^{1},q_{t}^{2}\right)
\end{array}\right)
\]
and 
\[
\left(\begin{array}{c}
q_{t+1}^{2}\\
p_{t+1}^{2}
\end{array}\right)=\mathcal{M}_{2}\left(\begin{array}{c}
q_{t}^{2}\\
p_{t}^{2}+\epsilon\left(q_{t}^{2}\right)+\kappa\left(q_{t}^{1},q_{t}^{2}\right)
\end{array}\right).
\]
The coupling between both maps is chosen to be  
$
\kappa (q_{t}^{1},q_{t}^{2})=-(K_{c}/2\pi)\sin(2\pi q_{t}^{1}+2\pi q_{t}^{2})
$.
The corresponding two degrees of freedom quantum evolution is given by 
the tensor product of the one degree of freedom maps, a $N^{2}\times N^{2}$ unitary matrix 
$
U_{j_{1}j_{2},k_{1}k_{2}}^{2D}=U_{j_{1}k_{1}}U_{j_{2}k_{2}}C_{j_{1}j_{2}},
$
with the coupling matrix (diagonal in the coordinate representation)
\[
C_{j_{1}j_{2}}=\exp\left\{ \left(\frac{iNK_{c}}{2\pi}\right)
\cos\left[\frac{2\pi}{N}\left(j_{1}+j_{2}\right)\right]\right\}, 
\]
where $j_{1},j_{2},k_{1},k_{2}\in\{0,\ldots,N-1\}$. We fix $K=0.25$ and $K_c=0.5$ (Anosov condition \cite{Ozorio 1994}), 
and $N=2^6$.
    
\begin{figure}[htp]
\hspace{0cm} \includegraphics[width=8cm]{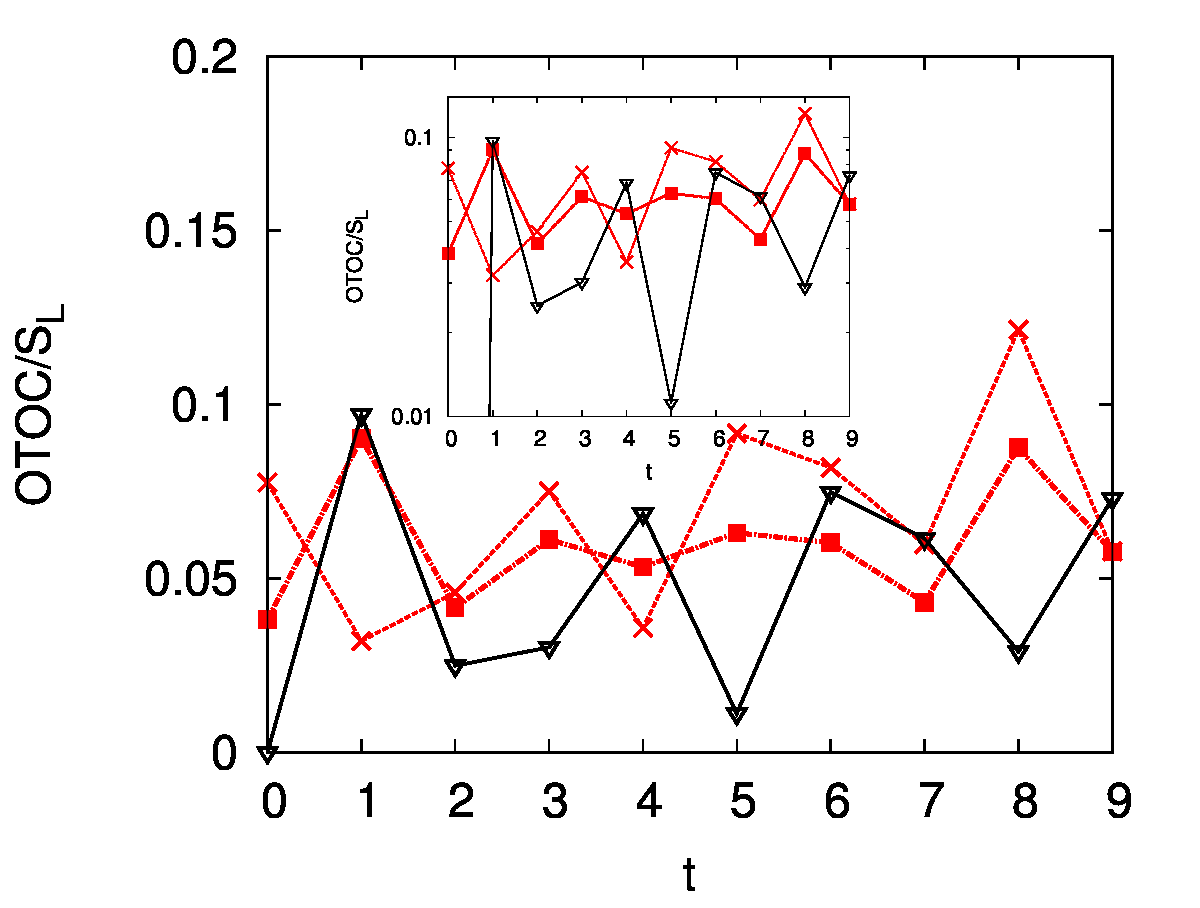} 
	\caption{(Color online) Evolution of the linear entropy $S_L$ ((black) solid line with down triangles), 
	and rescaled OTOCs for $\hat{A}=\hat{X}^{\rm 2D}$,  with 
	$\hat{B}=\hat{P}^{\rm 2D}$ ((red) dashed line with crosses), and $\hat{B}=\hat{\rho}(0)$ 
	((red) dot-dashed line with squares), as a function 
	of time $t$ (map time steps). $N=2^6$, EE maps case with initial coherent state centered at $(q,p)=(0.5,0.5)$. 
	Inset: log-linear version.}
	\label{fig1}
\end{figure}

{\em Results.}--In order to reach our twofold objective we have investigated the evolution of OTOCs much in the same way we have 
done in \cite{Bergamasco}, that is we have evaluated their behavior for 3 different dynamics. First, we 
consider the EE case. The initial (pure) state is constructed by placing a coherent state on each tori, 
both centered at $(q,p)=(0.5,0.5)$. This is a fixed point of the hyperbolic and elliptic maps. 
In Fig. \ref{fig1} we show the evolution of two OTOCs for $\hat{A}=\hat{X}^{\rm 2D}$,  having 
$\hat{B}=\hat{P}^{\rm 2D}$ in one case, and $\hat{B}=\hat{\rho}(0)$ in the other, as a function 
of the map time steps. Also we can see the evolution of the linear entropy $S_{\rm L}=1 - {\rm Tr} \rho_{1,2}^2$, 
which is a linear approximation to the von Neumann entropy $S_{\rm VN}$. 
We clearly see the same qualitative behavior, reflecting the lack of complexity 
growth due to the nature of the dynamics of both maps. We observe the same small oscillations indicative of the 
rotation of the distributions which remain localized \cite{Bergamasco}. 
Both OTOCs have been rescaled in Fig. \ref{fig1} (and all subsequent Figures) in order to make a comparison. 
In the inset we display the log-linear version, where no initial 
exponential growth can be identified for this case. 

\begin{figure}[htp]
\hspace{0cm} \includegraphics[width=8cm]{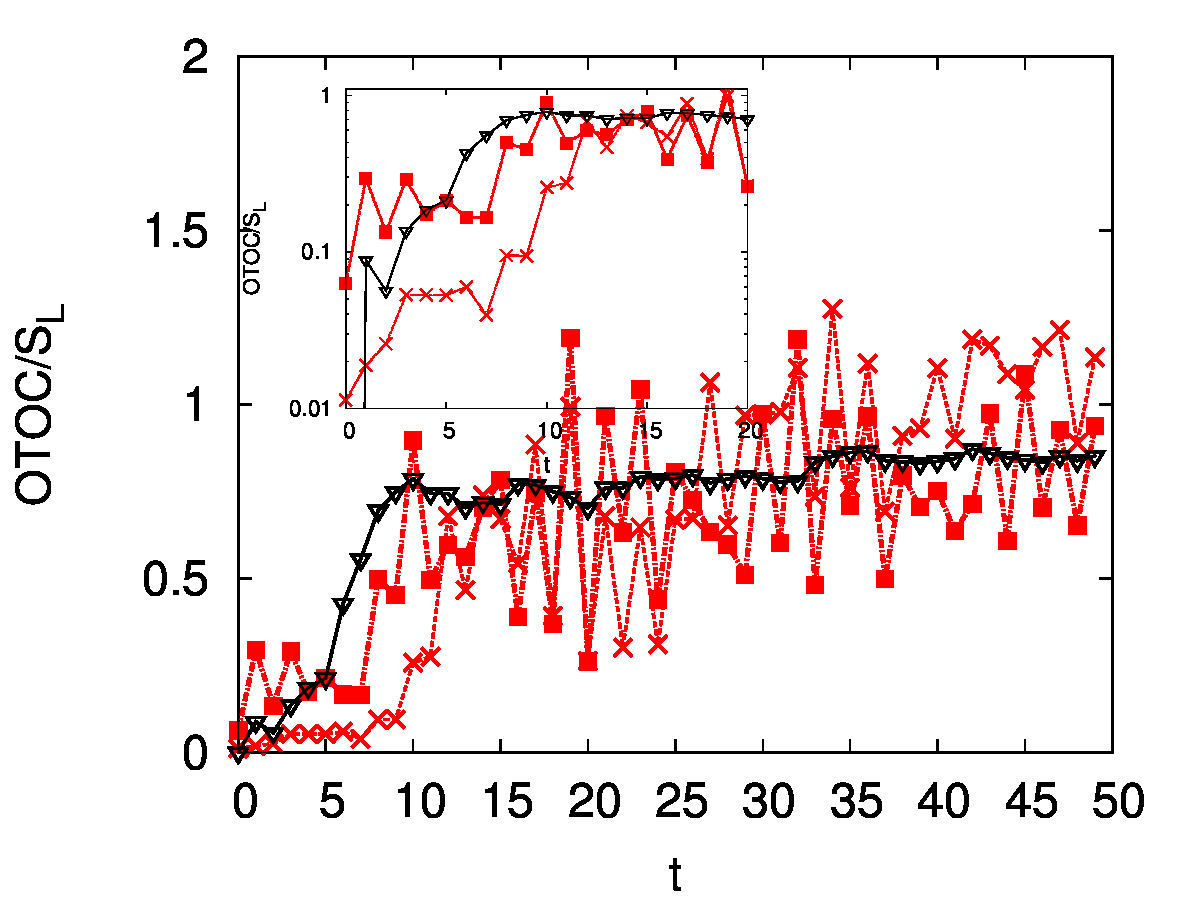} 
	\caption{(Color online) Evolution of the linear entropy $S_L$ ((black) solid line with down triangles), 
	and rescaled OTOCs for $\hat{A}=\hat{X}^{\rm 2D}$,  with 
	$\hat{B}=\hat{P}^{\rm 2D}$ ((red) dashed line with crosses), and $\hat{B}=\hat{\rho}(0)$ 
	((red) dot-dashed line with squares), as a function 
	of time $t$ (map time steps). $N=2^6$, EE maps case with initial coherent state centered at $(q,p)=(\pi/4,\pi/4)$. 
	Inset: log-linear version.}
	\label{fig2}
\end{figure}

It is interesting to see if the OTOC is able to detect the high sensitivity to the region of phase space at which 
the initial condition is located for the EE case, as we have previously seen by means of the WSE \cite{Bergamasco}. In fact, 
this is the case, and also the qualitative behavior is the same for the 3 quantities displayed in Fig. \ref{fig2}. 
There are clearly more fluctuations in the OTOCs, this will be explained later.
Moreover, in the inset it can be checked that no exponential growth is present. Despite this and fluctuations, the previously 
identified inflection point where quantum effects become important at $t \simeq 10$ \cite{Bergamasco} is roughly detected by the OTOCs.

\begin{figure}[htp]
\hspace{0cm} \includegraphics[width=8cm]{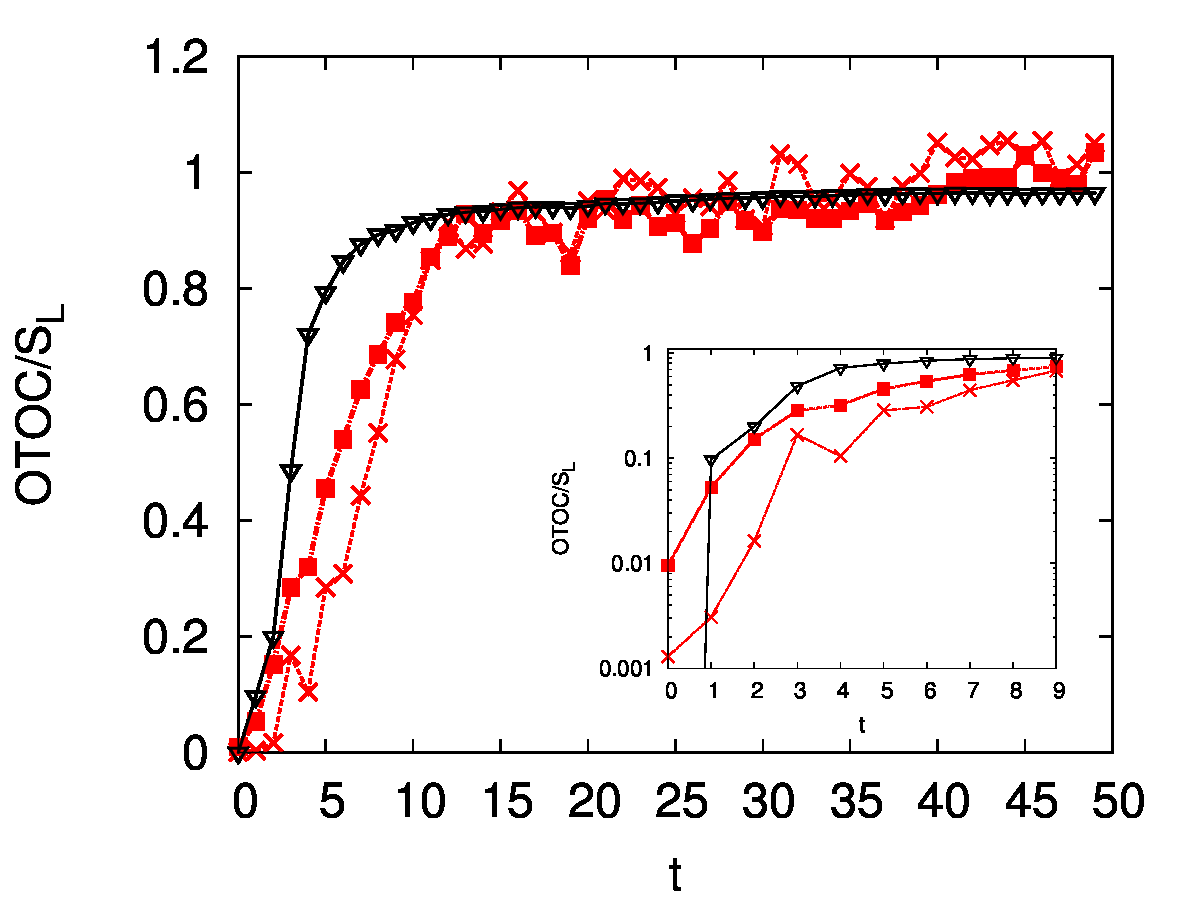} 
	\caption{(Color online) Evolution of the linear entropy $S_L$ ((black) solid line with down triangles), 
	and rescaled OTOCs for $\hat{A}=\hat{X}^{\rm 2D}$,  with 
	$\hat{B}=\hat{P}^{\rm 2D}$ ((red) dashed line with crosses), and $\hat{B}=\hat{\rho}(0)$ 
	((red) dot-dashed line with squares), as a function 
	of time $t$ (map time steps). $N=2^6$, HE maps case with initial coherent state centered at $(q,p)=(0.5,0.5)$. 
	Inset: log-linear version.}
	\label{fig3}
\end{figure}

We continue with the HE map case shown in Fig. \ref{fig3}, which again shows a good qualitative agreement between 
the linear entropy and the OTOCs behavior. The growth is slower for the correlators at early times, resembling more 
the von Neumann case which we will see in the following. Looking at the inset we cannot clearly identify an initial 
exponential growth of the correlators. Nevertheless the saturation behavior is very similar and this 
shows that the OTOC detects the main feature of the mixed dynamics scenario that we have already seen with the WSE: 
just one hyperbolic degree of freedom suffices to reach maximum complexity (this is accomplished for $t \simeq 200$, 
not shown here).

Finally, we turn to analyze the HH case, which has also been considered in \cite{Lakshminarayan2} very recently. 
Again, the agreement between OTOCs and $S_{\rm L}$ is remarkable, the $\hat{B}=\hat{\rho}(0)$ case being extremely 
good. If we look at the inset we can identify an initial exponential growth in full coincidence with previous 
studies. 
\begin{figure}[htp]
\hspace{0cm} \includegraphics[width=8cm]{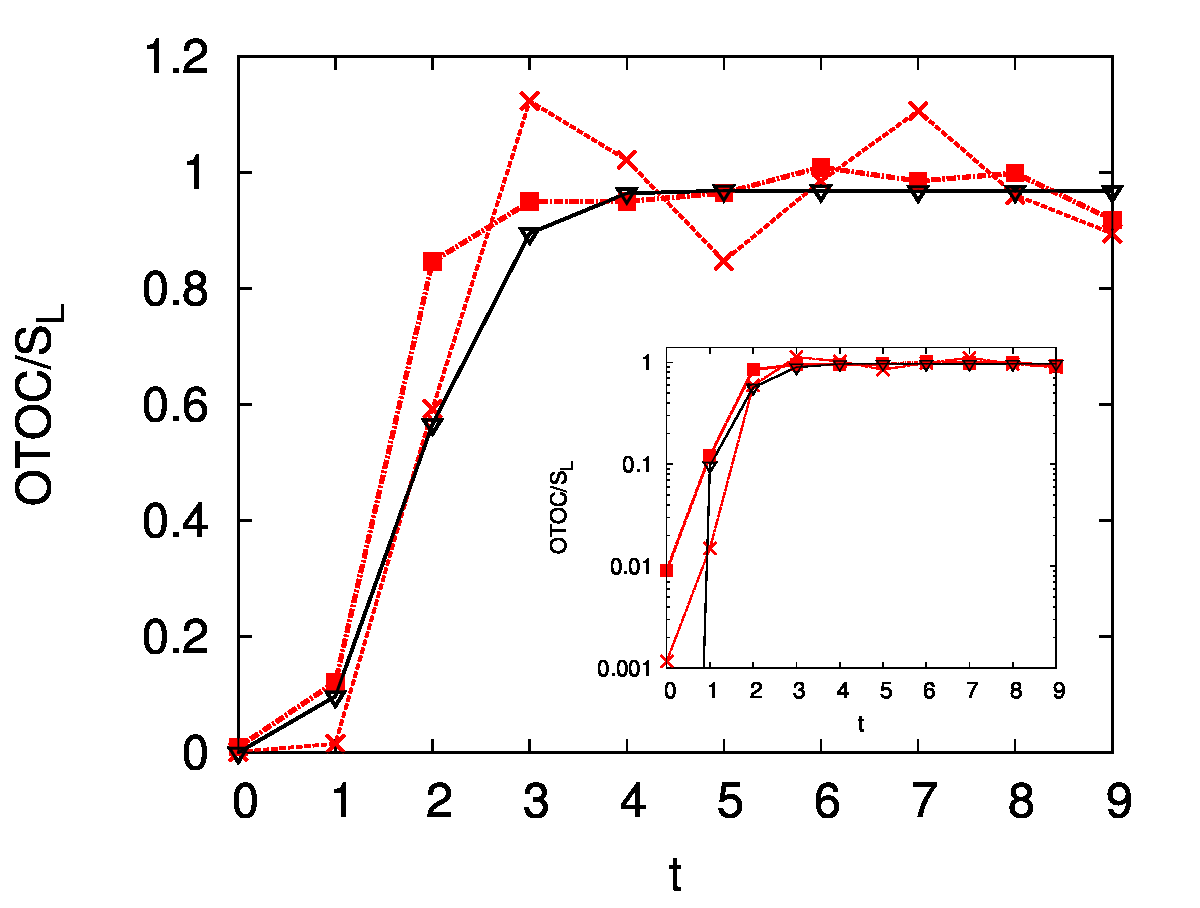} 
	\caption{(Color online) Evolution of the linear entropy $S_L$ ((black) solid line with down triangles), 
	and rescaled OTOCs for $\hat{A}=\hat{X}^{\rm 2D}$,  with 
	$\hat{B}=\hat{P}^{\rm 2D}$ ((red) dashed line with crosses), and $\hat{B}=\hat{\rho}(0)$ 
	((red) dot-dashed line with squares), as a function 
	of time $t$ (map time steps). $N=2^6$, HH maps case with initial coherent state centered at $(q,p)=(0.5,0.5)$. 
	Inset: log-linear version.}
	\label{fig4}
\end{figure}

But how can we explain this striking similarity between two seemingly 
different quantities, one coming from a global phase space analysis, and the other being a correlation 
related to a semiclassical interpretation based on local dynamics? An answer comes from the so called OTOC-RE theorem \cite{Hosur,Fan}. 
It establishes an equivalence between a sum of OTOCs (in fact, the sum of the 4 point correlator in which the OTOC 
can be split when the operators are Hermitian) over a complete basis of one of the subsystems (the operator that does not 
evolve is taken to be the initial state $\hat{\rho}(0)$), and the exponential of the second Renyi entropy. 
This result is usually expressed in the following shape: 
\begin {equation}
{\rm exp}{-S_1^{('2')}}=\sum_{\hat{M} \in 2} \left\langle \hat{M}(t) \hat{\rho}(0) \hat{M}(t) \hat{\rho}(0) \right\rangle
\end {equation}
where $S_1^{('2')}=-\log {\rm Tr}_1 \hat{\rho}_1^2$ is the second Renyi entropy, the sum runs over a complete basis of subspace $2$, 
and the usual thermal average is performed. 
It is easy to see that ${\rm exp}{-S_1^{('2')}}={\rm Tr} \rho_{1}^2$. 
In the OTOC Eq. \ref{eq:otocsplit}, we have a 2 point correlator term that 
in general can be considered to be constant (see for example \cite{Garcia, Lakshminarayan, Lakshminarayan2}, for chaotic cases). 
Hence, the OTOC becomes approximately proportional to $1 - {\rm Tr} \rho_1^2$, which is $S_{\rm L}$. 
Typically there are more oscillations in the 
OTOCs than in the linear entropy since we only consider one operator $\hat{A}(t)$ that belongs to both subspaces 
and not the complete basis of one of them as the OTOC-RE theorem prescribes. 
The equivalence expressed in this theorem is an indicator of an
average behavior of which our calculation is a fluctuation.

\begin{figure}[htp]
\hspace{0cm} \includegraphics[width=8cm]{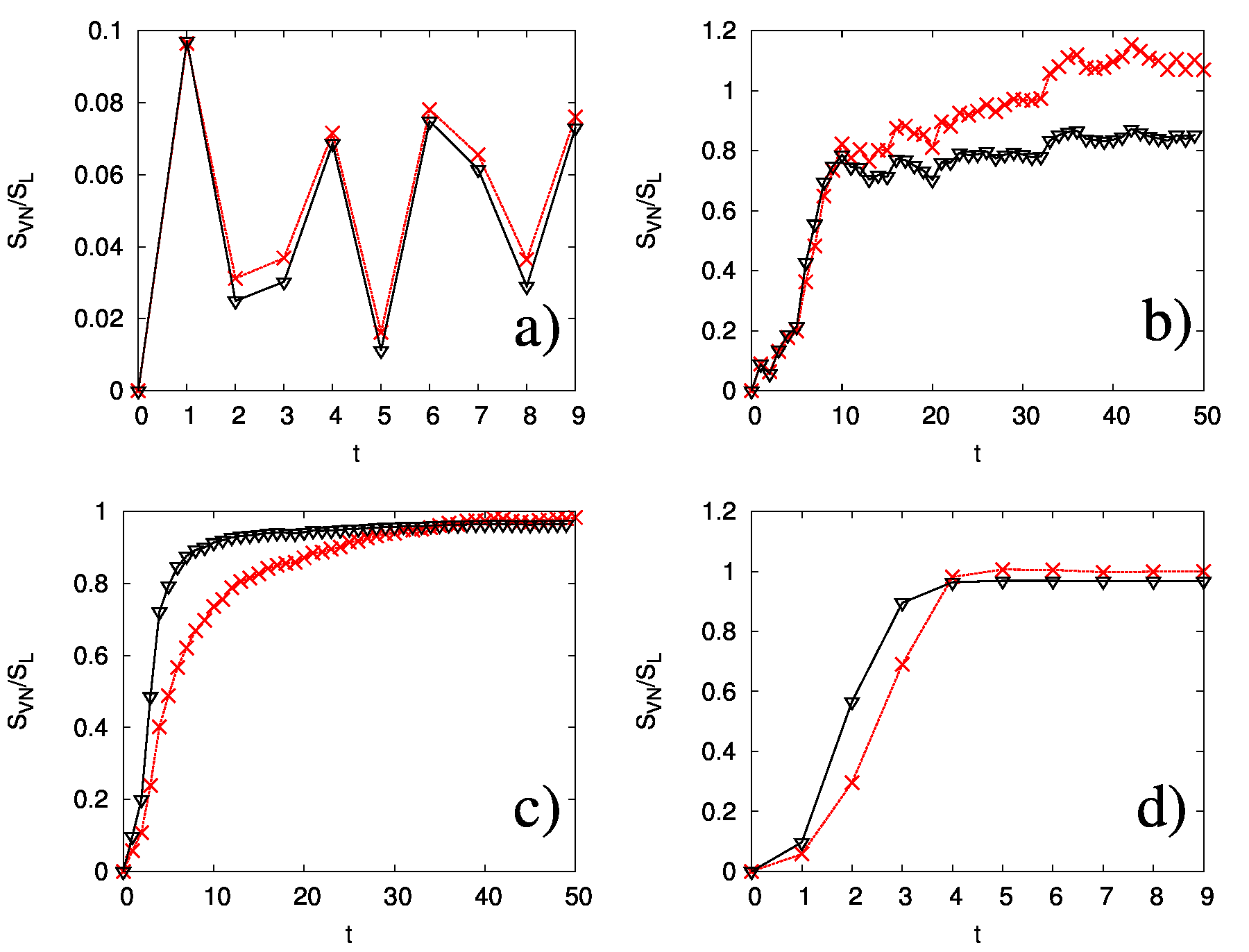} 
	\caption{(Color online) Comparison of rescaled von Neumann entropy $S_{\rm VN}$ ((red) dashed line with crosses) 
	with the linear entropy $S_{\rm L}$ ((black) solid line with down triangles) behavior as 
	a function of time $t$ (map time steps). In a) we show the EE case with initial coherent state 
	at $(q,p)=(0.5,0.5)$, in b) the EE case with $(q,p)=(\pi/4,\pi/4)$, in c) 
	the HE case with $(q,p)=(0.5,0.5)$, and in d) the HH case with $(q,p)=(0.5,0.5)$. In all panels $N=2^6$.}
	\label{fig5}
\end{figure}

In order to establish a complete link between the OTOCs and the WSE we show the rescaled von Neumann entropy evolution 
for the previous 4 cases, together with the linear entropy. We have rescaled $S_{\rm VN}$ in order to better compare it with 
$S_{\rm L}$ (in the last 2 cases we use the RMT saturation value \cite{Lakshminarayan3}). 
It becomes clear that $S_{\rm L}$  behaves much in the same way as $S_{\rm VN}$, despite 
being a linear approximation. It is worth mentioning that for the HE case (Fig. \ref{fig5} c)) the OTOC initial growth 
is closer to $S_{\rm VN}$. This will be investigated in the future.

{\em Conclusions}--Interest in OTOCs has grown very fast in the last couple of years, mainly motivated by their power to 
characterize quantum chaotic manifestations that could have important consequences in many body and high 
energy physics. In turn, the quantum chaos community is looking at its previous contributions from 
a new point of view. A third component comes from information theory which 
has established a precise connection between OTOCs and the second Renyi entropy via the OTOC-RE theorem. 

On the other hand, much has been done in one degree of freedom systems regarding OTOC measures 
but less is known in multi-partite cases. We have investigated a bi-partite system consisting of 
two coupled and perturbed cat maps with different dynamical scenarios, them being regular and chaotic. 
In all cases we have found that the behavior of OTOCs (semiclassically related to {\em local} measures 
of chaos) is qualitatively similar to that of the WSE. 
This latter is a complexity measure defined {\em globally} in phase space that treats the quantum and 
the classical distributions in the same way. This connection is formally explained by means 
of the equivalence between von Neumann/linear entropy and OTOCs when one of the operators 
considered is the initial density matrix. By choosing a pure initial state the WSE can be identified 
with the entropies. On the other hand, the OTOC-RE theorem describes the average behavior of the correlators 
for any choice of the evolving operator. In fact, even when considering the canonical $\hat{P}$ operator as the 
non evolving one, the agreement is very good, allowing to generalize this connection. 

It is worth mentioning that in many body systems the generic scenario involves chaotic and regular components. 
We have seen that one chaotic degree of freedom is enough for the complexity
measures to reach their maximum prescribed by RMT.
However, exponential growth of the OTOCs for localized initial conditions is absent if there is one
regular degree of freedom. In \cite{Maldacena} a bound is set for the OTOC Lyapunov exponential growth in  black
holes. In our examples we observe that any symmetry (constant of the motion) implies non exponential growth for the entropy. 
The consequences of this should be carefully explored. 

In the future we will investigate the OTOC-RE theorem and WSE connection in order to formalize it for 
generic initial states and operators. Different symmetry groups will be considered to obtain predictions on specific systems.

Support from CONICET is gratefully acknowledged.

\vspace{3pc}


\end{document}